\documentclass[smallextended]{svjour3}
\usepackage{amsmath,amssymb}
\usepackage{bm}
\RequirePackage{graphicx}
\RequirePackage{mathptmx}
\RequirePackage{latexsym}
\RequirePackage[numbers,sort&compress]{natbib}
\RequirePackage[colorlinks,citecolor=blue,urlcolor=blue,linkcolor=blue]{hyperref}

\begin{document}

\title{Evolution of density perturbations in fractional cosmology}

\author{S. M. M. Rasouli}

\institute{
S. M. M. Rasouli \at
Departamento de F\'{\i}sica,\\
Centro de Matem\'{a}tica e Aplica\c{c}\~oes (CMA-UBI),\\
Universidade da Beira Interior,\\
Rua Marqu\^es d'\'Avila e Bolama,\\
6200-001 Covilh\~a, Portugal\\
\email{mrasouli@ubi.pt}
}

\authorrunning{S. M. M. Rasouli}
\date{\today}
\maketitle

\begin{abstract}
 
We investigate the evolution of matter density perturbations within a fractional cosmological framework inspired by fractal space-time constructions in field theory, where a deformation of the integration measure induces non-locality and memory effects in the dynamics. Working in the matter-dominated era and 
adopting a covariant fluid-flow approach, we derive the modified growth equation for the density contrast and obtain exact analytical solutions.
The resulting dynamics depends explicitly on the fractional parameter $\alpha$ and smoothly reduces to the corresponding standard case in the limit $\alpha=1$. We show that the model admits both growing and decaying modes, and we identify the parameter range in which structure formation is physically viable. Focusing on the growing mode, we compute the evolution of density fluctuations from recombination to the present epoch. By confronting theoretical predictions with observational constraints from large-scale structure, in particular the $\sigma_8$ normalization and the Sachs-Wolfe effect, we derive a stringent upper bound on the fractional parameter, $\alpha\lesssim1.07$, which significantly improves upon previous constraints obtained at the background level. Our results show that the growth of density perturbations exhibits distinct fractional signatures, providing a sensitive observational probe of the underlying framework.
\end{abstract}
\keywords{Fractional cosmology \and Linear perturbations \and Structure formation
\and measure-induced deformation 
\and Exact solutions }

\maketitle

\section{Introduction}
\label{Intro}

Modern cosmological models, which are predominantly constructed under the assumptions of homogeneity and isotropy, provide an accurate description of the averaged evolution of the universe employing the standard or modified models \cite{rasouli2016exact,rasouli2019kinetic,rasouli2021geodesic,rasouli2022noncommutativity,rasouli2022geodesic,ildes2023analytic} in four or higher-dimensions \cite{doroud2009class,reyes2018emergence,mousavi2022cosmological,rasouli2022noncompactified,sania2023cosmic}. Nevertheless, background solutions alone are insufficient to explain the origin and evolution of realistic cosmic structures, such as the anisotropies of the cosmic microwave background (CMB) and the formation of galaxies and galaxy clusters. Consequently, the construction of a perturbation theory constitutes a fundamental pillar of any cosmological model. Within this framework, one can investigate the stability of the model, distinguish between growing and decaying modes, discriminate adiabatic from isocurvature perturbations, and compute both the primordial and evolved power spectra of cosmic structures.
Therefore, in this work, we investigate first order cosmological perturbations within a fractional cosmological framework, which, to the best of our knowledge, has not yet been systematically studied.

Several well-established and widely used frameworks for cosmological perturbation theory exist in the literature. In what follows, we briefly review the most relevant approaches and highlight their conceptual features, see for instance \cite{bardeen1980gauge, kodama1984cosmological,lifshitz1992gravitational,mukhanov1992theory,durrer20042,malik2009cosmological} and references therein.

\begin{description}
     
\item[Metric Perturbation Theory:] This framework was originally introduced by Lifshitz and later developed into a systematic and gauge-consistent formalism by Bardeen. In this approach, the spacetime metric and the energy--momentum tensor are perturbed around a Friedmann--Robertson--Walker (FRW) background, and the resulting linearized Einstein equations are solved. This framework naturally allows for the decomposition of perturbations into scalar, vector, and tensor modes, and it forms the backbone of precise calculations of CMB anisotropies and large-scale structure. 

Despite its generality, metric perturbation theory suffers from a strong dependence on the choice of gauge. In particular, the physical interpretation of certain perturbation variables, especially in specific gauges such as the synchronous gauge, may become obscure or ambiguous.

\item[Gauge-invariant formalism:] This formalism was developed to eliminate ambiguities associated with gauge choices in metric perturbation theory. In this framework, specific combinations of metric and matter perturbations are constructed such that they remain invariant under first-order coordinate transformations. These gauge-invariant variables play a central role in inflationary cosmology and in the interpretation of super-horizon perturbations.

Conceptually, this formalism is more transparent than gauge-fixed metric perturbations. However, from a fundamental standpoint, it does not constitute an independent framework, but rather an optimized reformulation of metric perturbation theory.

\item[Fluid-flow (covariant) approach:] An alternative and physically intuitive framework is provided by the fluid-flow, or covariant, approach. This method was first proposed by Hawking \cite{hawking1966perturbations} and subsequently developed by several authors \cite{lyth1990evolution}. Instead of perturbing the spacetime metric, one focuses on the local fluid equations (such as the continuity equation and the Raychaudhuri equation) along comoving worldlines. Within this approach, the evolution equation for density perturbations can be derived directly.

Using this framework, one can recover Bardeen's equation for density perturbations without explicit reference to metric perturbations, and further generalize it to multi-fluid systems, leading naturally to the Kodama--Sasaki equations \cite{kodama1984cosmological,lyth1990evolution}. It has been believed that the fluid-flow approach is often simpler and more physically transparent, while exhibiting a clear structural analogy with Newtonian perturbation theory.

\item[Action-based and Hamiltonian approaches:] In certain applications-notably in the quantization of perturbations and in the analysis of fundamental stability conditions, cosmological perturbations are studied by expanding the action to second order in perturbation variables. This approach is particularly powerful in inflationary cosmology, where it provides a direct route to the quantization of scalar and tensor modes. However, in effective or generalized models, the technical implementation of this method may become highly non-trivial.
\end{description}

One or more of the mentioned perturbative approaches, chosen according to their efficiency and suitability, have been applied in Newtonian cosmology (NC), the standard cosmological framework of general relativity, as well as in various modified cosmological models. In particular, in standard NC, perturbations are obtained using the continuity, Euler, and Poisson equations. The fundamental perturbation variable is the density contrast, whose evolution equation governs the growth of cosmic structures. This formulation represents the Newtonian counterpart of the fluid-flow approach \cite{lyth1990evolution} in relativistic cosmology. Accordingly, in this work, we show that, within modified Newtonian cosmological models, such as the emergent $\Lambda$CDM cosmology \cite{rasouli2026emergent}, constructed inspired by the frameworks developed in \cite{calcagni2010quantum,calcagni2010fractal}, the adoption of fluid-flow approach constitutes a natural and effective approach for the analysis of cosmological perturbations.

In recent years, various approaches based on fractional calculus have been employed to modify cosmological models at both the classical and quantum levels \cite{jalalzadeh2022sitter,varao2024fractional,marroquin2024conformal,junior2025fractional,canedo2025quantum,da2025fractional,vacaru2012fractional,calcagni2021classical,rasouli2022inflation, micolta2025fractional,Rasouli:2025qix,calcagni2010quantum,el2026fractional}.
Among them, one of the interesting approaches typically extends standard dynamical structures, for instance, through modifications of the integration measure \cite{calcagni2010quantum,calcagni2010fractal} or the introduction of time-dependent kernels \cite{el2017fractional,micolta2023revisiting,rasouli2024fractional,el2013non,Rasouli:2025qix}, leading to the emergence of non-local behavior and memory effects in the equations of motion. Among these developments, multifractional theories \cite{calcagni2021multifractional} provide a prominent example, where a deformation of the spacetime measure induces controlled deviations from standard dynamics.

Motivated by the mentioned fractional implications, in Ref.~\cite{rasouli2026emergent} we proposed a minimal and tractable fractional model, in which a time-dependent kernel is incorporated at the level of the action, resulting in an effective modification of the dynamics. This framework preserves a well-defined classical limit and remains compatible with general relativity in the weak-field regime \cite{rasouli2026minimal}, while at the cosmological level it naturally reproduces the standard relativistic cosmological equations in an emergent manner \cite{rasouli2026emergent,rasouli2026inflation}.

Despite the success of the fractional models in describing the background evolution and providing a unified account of different cosmic epochs \cite{el2012gravitons,rami2015fractional,shchigolev2016testing, rasouli2021broadening,socorro2023anisotropic}, a systematic analysis of cosmological perturbations in such fractional frameworks is still largely lacking. We note that perturbations in a fractional scalar-field cosmological model have recently been investigated in a fully general setting \cite{Rasouli:2025qix}; however, explicit cosmological solutions were not obtained in that context. Moreover, the fractional cosmological model considered here differs substantially, both conceptually and structurally, from scalar-field-based constructions. For this reason, the main goal of the present work is to investigate the dynamics of density perturbations and the growth of structures within the framework of emergent fractional cosmology and to explore their observational implications.

The main objective of the present work is to investigate the implications of the fractional framework established in \cite{rasouli2026emergent} at the level of cosmological perturbations. Concretely, we focus
on the following aspects: (i) the derivation of the fractional hydrodynamical equations, namely the continuity, Euler, and Poisson equations;  
(ii) the construction of the corresponding linear perturbation equations and,
in particular, the resulting fractional density-growth equation;  
(iii) analytical solution of the density-growth equation; and  
(iv) the extraction of an upper observational bound on the fractional
parameter $\alpha$ using recent cosmological data, which further tightens the
theoretical and observational constraints inherent to the model in the background level.

The structure of this paper is organized as follows. 
In the next section, we briefly
review the \textit{emergent relativistic cosmology} from a minimal fractional extension of the classical action \cite{rasouli2026emergent} and outline the corresponding consequences. 
In Section \ref{Pert}, we formulate fractional hydrodynamical equations. By introducing linear (first-order) perturbations to these equations, we derive the matter density growth equation and solve it exactly. We then compare the fractional results with their standard Newtonian and relativistic counterparts. Finally, using observational data, and in particular the Sachs-Wolfe relation, we interpret the role of the fractional parameter $\alpha$ and derive an upper bound on it. Section \ref{Concl} summarizes the results obtained in this paper and provides a foundation for further developments to be presented in future studies.

\section{\texorpdfstring{Emergent $\Lambda$CDM cosmology and background solutions}{Emergent LambdaCDM cosmology and background solutions}}
\label{Frac-New-Mech}

In Ref. \cite{rasouli2026emergent}, we established a fractional framework, which can be regarded as a minimal and reduced realization of the quantum field theory on fractal spacetime \cite{calcagni2010fractal,calcagni2010quantum}. In the latter framework, the standard integration measure in the action is replaced by a non-trivial Stieltjes measure that effectively characterizes fractal features of spacetime. This modification naturally leads to anomalous scaling behavior, non-locality, and memory effects at the level of the dynamical equations.

Moreover, we have shown that the resulting fractional cosmological equations reproduce, at the background level, the standard relativistic cosmological equations. From this perspective, the model can be viewed as an emergent cosmological framework that establishes a conceptual bridge between Newtonian and relativistic cosmology. In particular, in the limit $\alpha=1$, all fractional modifications vanish and the model consistently reduces to its standard Newtonian counterpart, for a detailed study, see \cite{rasouli2026emergent}.

An important feature of this fractional framework is that it enables the description of cosmological epochs, such as the radiation-dominated era, the late-time accelerated expansion, and even the early inflationary phase \cite{rasouli2026emergent,rasouli2026inflation}, that are not accessible within standard Newtonian cosmology. Furthermore, the fractional structure allows for a broader class of cosmological solutions compared to the conventional relativistic framework, while observational consistency imposes meaningful constraints on the allowed range of the fractional parameter $\alpha$.

In the following, we present a concise overview of the fractional Newtonian model, the corresponding emergent relativistic cosmological framework, and the exact solutions derived within this context; for more comprehensive discussions see Refs. \cite{rasouli2026emergent,rasouli2026inflation,rasouli2026minimal}. 

In our previous work \cite{rasouli2026emergent}, inspired by \cite{calcagni2010fractal,calcagni2010quantum}, we introduced a modified classical (Newtonian) action as

\begin{eqnarray}\label{New-fr-action}
 S_{\alpha}=
 \!\!\frac{1}{\Gamma(\alpha)}\int_{0}^{\bar t} \xi({\tau} )
 \left(T-V_{\rm eff}\right)\,d{\tau},
\end{eqnarray}
with
\begin{equation}\label{L-xi}
 \xi({\tau})\equiv\left(\frac{{\bar t} - \tau}{\tau_0}
\right)^{\alpha-1},
\end{equation}
where $\tau_0>0$ is a fixed reference time scale included solely to make the time-dependent kernel $ \xi$ dimensionless, $\alpha$ is the fractional parameter, $T=\frac{1}{2}\,m (\frac{d {\mathbf{r}}}{d\tau} )^2$ is the standard kinetic energy, and
$V_{\rm eff}=V_{\rm eff}(\mathbf r;\alpha)$ represents an effective potential which depends explicitly on the fractional parameter $\alpha$, such that in the limit $\alpha=1$, it reduces to the standard potential $V(\mathbf{r})$ associated with conservative forces.


It is worth emphasizing that the fractional action introduced in~\eqref{New-fr-action} is defined through a deformation of the integration measure, rather than through fractional derivatives or through a prescribed memory kernel at the level of the equations of motion. More precisely, the present construction may be regarded as a minimal effective realization of the ideas underlying fractional action principles and Stieltjes-type measures used in fractional and multifractional spacetime models \cite{calcagni2010fractal,calcagni2010quantum}. While more general modified-measure frameworks may involve additional geometric structures or effective degrees of freedom, the present model introduces only a fractional time-dependent kernel in the action measure and does not add any independent dynamical degree of freedom. The parameter $\alpha$ therefore characterizes the effective temporal measure, and the standard Newtonian action is continuously recovered in the limit $\alpha=1$.
This should also be distinguished from alternative formulations of fractional or non-Markovian dynamics based on memory kernels or Onsager-Machlup type effective actions. The present approach follows a different route: it uses a minimal fractional deformation of the action measure as an effective description, motivated by fractional-measure constructions, in order to test whether standard cosmological dynamics and its perturbative sector can be reproduced without enlarging the dynamical field content.
A more detailed discussion of the physical and philosophical motivations underlying this action, as well as its relation to other fractional frameworks, is given in Ref. \cite{rasouli2026emergent}. Here we have summarized only the points needed to make the present analysis self-contained.

\begin{itemize}
\item
Employing transformation $\bar{t}-\tau\equiv t$, the modified equations from the variation of the action~\eqref{New-fr-action} take the form
\begin{eqnarray}
m\,\ddot{\mathbf{r}}
= -\nabla V_{\rm eff}(\mathbf{r})- m \gamma_\alpha(t) \,\dot{\mathbf{r}}, \hspace{5mm} \gamma_\alpha(t)\equiv \frac{(\alpha - 1)}{t},
\label{fr-eq-3d}
\end{eqnarray}
where $\gamma_\alpha(t)\equiv  \frac{\dot{\xi}}{\xi}$, $\nabla$ denotes the gradient operator with respect to $\mathbf r$.

\item
From equation \eqref{fr-eq-3d}, one can show that an effective mechanical energy can still
be defined, which remains conserved in the absence of external forces,
\begin{eqnarray}\label{Fr-cons}
\mathcal{E^{^{\rm mech}}_{\rm eff}}\equiv T + V_{\rm eff} +T_\alpha = \text{constant},
\end{eqnarray}
where 
\begin{eqnarray}\label{frac-T}
T_\alpha \equiv  \, 2( \alpha-1) \int_{t_i}^{t_f} T(t)\, d\ln t,
\end{eqnarray}
can be interpreted as the fractional kinetic energy, which corresponds to a time-averaged version of its standard counterpart.

\item
In a particular case, when the standard and fractional contributions to the potential are separable, namely $V_{\rm eff}(\mathbf r;\alpha)=V(\mathbf r)+V_{\alpha}(\mathbf r;\alpha)$,
the model admits a simpler and more transparent interpretation. In this case,
the quantity $E_\alpha \equiv T_\alpha + V_\alpha$
can be naturally interpreted as the fractional mechanical energy of the system \cite{rasouli2026emergent}.

\item
In the particular limit $\alpha=1$, all fractional quantities reduce smoothly
to their standard Newtonian counterparts. Consequently, the classical form of
Newton's second law, the kinetic and potential energies, and the standard 
mechanical energy are exactly recovered, ensuring a well-defined classical
limit of the theory.
\end{itemize}

In what follows, we briefly review the fractional cosmological model emerging from this modified dynamics.

We consider a homogeneous and isotropic spherical region of physical radius $a(t)$, filled with pressure-less matter (dust) of uniform density $\rho_m$. A test particle of unit mass ($m=1$) is assumed to lie on the boundary of the sphere.
In standard NC, assuming that the dust matter is governed by the Newtonian gravitational potential $\Phi_{\rm N}$, the gradient of the potential inside a homogeneous sphere takes the well-known form
\begin{equation}\label{N-Pot}
    \nabla_{\!r}\,\Phi_{\rm N}=\frac{4 \pi G}{3} \rho_m \mathbf{r},
\end{equation} 
where $\mathbf r$ is the position vector measured from the center of the sphere. Equation \eqref{N-Pot} together with the Euler and continuity equations, yields the cosmological equations associated with the standard NC.

However, within the emergent $\Lambda$CDM model, the
dynamics is enriched by additional fundamental ingredients that originate from fractional modification, so it reduces to the standard action in the limit $\alpha=1$. Remarkably, their combined effect allows the
cosmological dynamics to remain fully self-consistent, while the
cosmological equations corresponding to the $\Lambda$CDM model naturally
emerge from the underlying fractional dynamics \cite{rasouli2026emergent}.

Let us summarize the set of equations associated with the emergent $\Lambda$CDM model. 
For the sake of simplicity, we assume that
\begin{eqnarray}\label{eff-pot-cos}
\Phi_{\rm eff}=\Phi_{_{\rm N}}+\Phi_{\alpha},
\end{eqnarray}
where $\Phi_{\alpha}$ is the fractional contribution of the gravitational potential, which is assumed to vanish in the limit $\alpha=1$.
Moreover, we use the following definitions
\begin{eqnarray}\label{cos-T-def}
T&\equiv& \frac{1}{2}\dot{a}^2, \hspace{10mm}
T_{\alpha}\equiv
2(\alpha-1)\int_{t_i}^{t_f}T(t)\, d\ln t,\\\nonumber\\
\label{cos-E-def}
E&\equiv& T+\Phi_{_{N}}, \hspace{5mm} E_{\alpha}\equiv T_{\alpha}+\Phi_{\alpha}.
\end{eqnarray}

With the above definitions, and by applying the fractional dynamics, it has been shown that the equations of the emergent $\Lambda$CDM model take the form \cite{rasouli2026emergent}
\begin{eqnarray}\label{eff-frd}
 H^2=\frac{8 \pi G}{3}\rho_{_{\rm eff}}-\frac{\mathcal{K}}{a^2},
\end{eqnarray}
\begin{eqnarray}\label{acc-eff-eq}
\frac{ \ddot{a}}{a}=-\frac{4\pi\,G}{3}
\Big(\rho_{_{\rm eff}}+3 \,p_{_{\rm eff}}\Big),
\end{eqnarray}
\begin{eqnarray}\label{cons-eq-cos}
\dot{\rho}_{_{\rm eff}}+3H (\rho_{_{\rm eff}}+p_{_{\rm eff}})=0,
\end{eqnarray}
where $H\equiv\dot{a}/a$ is the Hubble parameter and 
\begin{eqnarray}\label{K-def}
\mathcal{K}&\equiv& -2 \mathcal{E}=-2(E+E_{\alpha})=\mathrm{constant}, \\
\label{rho-eff}
\rho_{_{\rm eff}}&\equiv&\rho_m+\rho_{\alpha}, \hspace{5mm}\rho_{\alpha}\equiv-\frac{3}{4 \pi G}\left(\frac{E_{\alpha}}{a^2}\right),
\end{eqnarray}
\begin{eqnarray}\label{p-eff}
p_{_{\rm eff}}=p_{\alpha}\equiv\frac{1}{4\pi\,G}\frac{1}{a^2}
\frac{d}{da}\Big(a\,E_{\alpha}\Big).
\end{eqnarray}

Equations \eqref{eff-frd}-\eqref{cons-eq-cos} constitute the fundamental equations of the emergent $\Lambda$CDM model and define a closed and self-consistent dynamical system.

We should note that the acceleration equation \eqref{acc-eff-eq} can be rewritten as
\begin{eqnarray}\label{acc-eq}
 \ddot{a}(t)+\gamma_{\alpha}(t)\,\dot{a}(t)+\frac{d\Phi_{\rm eff}(a)}{da}=0,
\end{eqnarray}
which will be referred to in the next sections.

In what follows, let us summarize the main features of the emergent $\Lambda$CDM cosmology.

\begin{itemize}

\item
In the special limit $\alpha = 1$, all fractional contributions are completely
switched off and the model reduces exactly to standard NC.

\item
In the general case $\alpha \neq 1$, equations \eqref{eff-frd}-\eqref{cons-eq-cos} become dynamically
equivalent to the cosmological equations of the $\Lambda$CDM model. In particular,
for $\mathcal{K}=0$, it has been shown that these equations are fully
self-consistent and are able to generate all background solutions associated
with the $\Lambda$CDM model. In practice, the reconstruction of the cosmic
evolution in each individual epoch naturally leads to the emergence of a
specific fractional potential $\Phi_\alpha$ characterizing that epoch.
Importantly, the explicit determination of these epoch-dependent fractional
potentials provides a valuable guideline for conjecturing a single unified
fractional potential capable of describing the entire cosmic history within a
single framework \cite{rasouli2026emergent,rasouli2026inflation}.

\item

Based on the definitions of the effective energy density \eqref{rho-eff} and the
effective pressure \eqref{p-eff}, one can show that these quantities satisfy a
well-defined continuity equation as
 \begin{eqnarray}\label{cons-eq-fr}
\dot{\rho}_{\alpha}+3H (\rho_{\alpha}+p_{\alpha})=0.
\end{eqnarray}
 Moreover, the continuity equations \eqref{cons-eq-cos} and \eqref{cons-eq-fr} naturally reproduce the standard continuity equation governing
ordinary matter in the standard NC:
\begin{eqnarray}\label{cons-eq-rho-m}
\dot{\rho}_{\rm m}+3H\rho_{\rm m}=0.
\end{eqnarray}
The fact that the continuity equations
are satisfied separately for the ordinary matter sector and the fractional
sector indicates that the two components are not directly coupled. This feature
represents a distinctive property of the emergent $\Lambda$CDM cosmology and leads to non-trivial physical consequences at both the background
and perturbative levels (see the next section).
\end{itemize}

In our previous works, we have shown that, by employing the fractional
cosmological equations, one can exactly obtain the background solutions
corresponding to dynamics of the very early universe, radiation-dominated, matter-dominated, and present
accelerating phases of the universe in agreement with observations \cite{rasouli2026emergent,rasouli2026inflation}.
This analysis naturally leads to the introduction of a fully general and
unified fractional potential capable of describing the entire cosmic history
within a single theoretical setup (see Refs. \cite{rasouli2026emergent,rasouli2026inflation}).

Motivated by this general potential, we subsequently proposed a simpler
interpolating effective potential (For a detailed study, see \cite{rasouli2026emergent}.):
\begin{eqnarray} \label{par-eff-pot}
\Phi_{_{\rm eff}}(a;\alpha)=\Phi_{\rm N}(a)+\Phi_{\alpha}(a;\alpha),
\end{eqnarray}
where $\Phi_{\alpha}(a;\alpha)$ is given by
\begin{eqnarray}  \label{par-phi-alpha}
\Phi_{\alpha}(a;\alpha)=R(\alpha)\,a^{-2}\,W_{\rm rad}(a)+L(\alpha)\,a^{2}\,W_{\Lambda}(a),
\end{eqnarray}
which is effectively switched on
and off between different cosmological epochs through suitable switching
functions 
\begin{equation}\label{switch}
W_{\rm rad}(a)
\equiv \frac{1}{1+\bigl(\tfrac{a}{a_{\rm eq}}\bigr)^{n_r}},
\qquad
W_{\Lambda}(a)
\equiv \frac{1}{1+\bigl(\tfrac{a_{\Lambda}}{a}\bigr)^{n_\Lambda}}.
\end{equation}
In equation \eqref{switch}, $n_r,n_\Lambda>0$ control the sharpness of the transitions, and
$a_{\rm eq}$, $a_{\Lambda}$ are the equality and acceleration
transition scales, respectively. Moreover, we consider a specific case $\mathcal{K}=0$, which corresponds with the spatially flat FLRW cosmology and is in agreement with the current cosmological observations \cite{Planck:2018vyg}.

In \eqref{par-phi-alpha}, since the fractional potential associated with the matter-dominated
era exhibits the same functional dependence as the Newtonian gravitational
potential, the fractional contribution $\Phi_\alpha$ can be consistently set
to zero in this regime without any loss of generality, and only the Newtonian
potential is retained.

Furthermore, in the radiation-dominated and late-time accelerating phases,
the Newtonian potential $\Phi_{\rm N}$ plays a subdominant role compared to the fractional
potential. Consequently, its contribution is neglected in these epochs.
Nevertheless, throughout all cosmological eras, the friction-like term
originating from the time kernel in the fractional action (i.e. the term including $\gamma_{\alpha}(t)$) is fully preserved and continues to influence the dynamics.

Under these assumptions, a set of analytically tractable and physically
interesting solutions emerges, allowing for a transparent and controlled
analysis of the role played by the fractional parameter $\alpha$.
It is worth emphasizing that all these solutions consistently satisfy the
fractional cosmological equations \eqref{eff-frd}-\eqref{cons-eq-rho-m},
ensuring the full dynamical self-consistency of the model.
For further details on the explicit form of the solutions, we refer the reader
to Ref.~ \cite{rasouli2026emergent}.

Without entering into the technical details, we summarize below the exact
solutions associated with the fractional potential \eqref{par-phi-alpha} obtained in
Ref.~\cite{rasouli2026emergent}.
Using equation \eqref{cons-eq-rho-m}, we obtain:
\begin{eqnarray}\label{dens-m}
\rho_{\rm m}(a)= \rho_i\left(\frac{a}{a_i}\right)^{-3}\equiv\tilde{\rho}_{\rm mi }\,a^{-3}, 
\end{eqnarray}
where $\tilde{\rho}_{mi }>0$ is an integration constant.

As mentioned, in the matter-dominated regime, without loss of generality, one can assume
$\Phi_{_{\rm eff}}\approx \Phi_{\rm N}$.

Therefore, we obtain
\begin{eqnarray}\label{mat-a-special}
a(t)&=& a_i \,t ^{\frac{2}{3}}, \hspace{15mm} a_i\equiv\left(\frac{6 \pi \,G \tilde{\rho}_{mi } }{4-3 \alpha}\right)^{\frac{1}{3}}\\\nonumber\\
\label{mat-rho-special}
\rho^{(m)}_{_{\rm eff}}&=&\frac{\tilde{\rho}_{mi }}{4-3 \alpha} \left(\frac{1}{a}\right)^{3}, \hspace{10mm} p^{(m)}_{_{\rm eff}}=0.
\end{eqnarray}

Here, we refrain from presenting the exact background solutions for the radiation-dominated era and the current accelerated phase; the reader is referred to Ref. \cite{rasouli2026emergent} for a detailed discussions. Nevertheless, it is worth noting that the simultaneous application of theoretical and observational constraints to the background solutions across all cosmological  epochs leads to the following common range for the fractional parameter
\begin{equation}\label{alpha-obs}
  0.8 \;\lesssim\; \alpha \;\lesssim\; 1.2.
\end{equation}

In the next section, we use the result \eqref{alpha-obs}, and the background solutions reviewed above to solve and analyze the corresponding density-growth equation.


\section{\texorpdfstring{Linear perturbation in emergent $\Lambda$CDM cosmology}{Linear perturbation in emergent Lambda CDM cosmology}}
\label{Pert}

In this section, we follow the same procedure as used in standard Newtonian cosmology  \cite{savedoff1962growth,adams1975exact,lyth1990evolution,arcuri1994growth,green2012newtonian,green2014well,fidler2017general} 
to derive the first-order perturbation equations in our fractional model.
Concretely, we formulate the fractional continuity, Poisson, and Euler
equations for the matter-dominated
regime, in which only the time-dependent kernel $\xi(t)$ is present. We then derive the corresponding first-order perturbation equations and
analyze their implications for the growth of matter density fluctuations.

\subsection{Fractional hydrodynamical equations: matter-dominated case study}
\label{hydro}

In proper coordinates $\mathbf r$, the  fractional hydrodynamical 
equations of the pressure-less fluid in our model are 
governed by the modified (fractional) continuity, Euler, and Poisson equations as
\begin{eqnarray}\label{con-eq-1}
\frac{\partial \rho_{_{\rm eff}}}{\partial t}+\nabla_{\!r}\!\cdot \bigl(\rho_{_{\rm eff}}\,\mathbf v\bigr) = 0,
  \end{eqnarray}
  \begin{eqnarray}   \label{Eul-eq}
  \frac{\partial \mathbf v}{\partial t} + (\mathbf v \cdot \nabla_{\!r}) \mathbf v
   + \gamma_{\alpha}(t)\,\mathbf v
   = - \nabla_{\!r} \Phi_{\rm N},
    \end{eqnarray}
    \begin{eqnarray}   \label{Poi-eq}
\nabla_{\!r}^{2} \Phi_{_{\rm N}}=4 \,\pi\, G \,\rho_{\rm m} ,
   \end{eqnarray}
where $\mathbf v(t,\mathbf r)$ is the velocity field.

For a homogeneous and isotropic background, we adopt the Hubble flow ansatz
$\bar{\mathbf v}(t,\mathbf r)=H(t)\, \mathbf{r}$.  Then, using equations~\eqref{con-eq-1}, \eqref{Eul-eq}, and \eqref{Poi-eq}, one can directly derive the system of equations
of our fractional model:
\begin{eqnarray}\label{con-back}
{\dot{\bar \rho}_{_{\rm eff}}} + 3 H \bar \rho_{_{\rm eff}} = 0,
\end{eqnarray}
\begin{eqnarray}\label{Eul-back}
\frac{\ddot a}{a} + \gamma_{\alpha} \,H
   = - \frac{4\pi G}{3}\,\bar\rho_{\rm m},
     \end{eqnarray}
\begin{eqnarray}\label{Poi-back}
\nabla_{\!r}^{2}\bar\Phi_{_{\rm N}} = 4 \,\pi\, G \,\bar\rho_{\rm m},
\end{eqnarray}
where the quantities with a bar denote their homogeneous
background value. It is observed that the set of equations \eqref{con-back}-\eqref{Poi-back} exactly coincides with the main system of equations of our fractional model, namely equations \eqref{cons-eq-cos} and \eqref{acc-eq}, in the particular case $\Phi_\alpha=0=p_{\alpha}$. 
It is clear that in the special case where $\alpha=1$,
all of the above equations reduce to their
standard Newtonian counterparts.

\subsection{Linear Perturbations}

Let us briefly outline the following procedure for obtaining the
first–order (linear) perturbation equations from the system \eqref{con-eq-1}-\eqref{Poi-eq}.

To obtain the perturbed equations, let us switch to 
the comoving coordinate $\mathbf x$ related to the proper coordinate  $\mathbf r$ as $\mathbf r = a(t)\,\mathbf x$, 
such that the corresponding spatial gradients are related by $\nabla_{\!r} = \frac{1}{a}\,\nabla_{\!x}$.
We perturb the velocity field, the density and the effective potential as 
\begin{eqnarray}\nonumber
\rho_{_{\rm eff}}(\mathbf x,t)&=&\bar\rho_{_{\rm eff}}(t)\,[1+\delta(\mathbf x,t)], \hspace{5mm}
\mathbf v = H a\, \mathbf x + \mathbf u,\\\nonumber\\
\Phi_{_{\rm N}}(\mathbf x,t)&=&\bar\Phi_{_{\rm N}}(t)+\phi(\mathbf x,t),
\label{pert-rel}
\end{eqnarray}
where $ u\ll v$ is the peculiar velocity and $\phi$ is the perturbation of the potential. We should note that, in the particular case where $\Phi_\alpha=0=p_\alpha$, we obtain a relation between the density contrasts as $\delta_{\rm eff}(\mathbf x,t)\equiv\displaystyle \frac{\rho_{_{\rm eff}}-\bar\rho_{_{\rm eff}}}{\bar\rho_{_{\rm eff}}}=\delta_{\rm m}\equiv\delta$, 

Substituting $\rho(\mathbf x,t)$ from equation \eqref{pert-rel} into the continuity equation \eqref{con-eq-1},
using the background relation \eqref{con-back},
we obtain the linear perturbed continuity equation:
\begin{equation}\label{Con-lin}
\dot\delta + \frac{1}{a}\,\nabla_{\!x}\!\cdot \mathbf u = 0,
\end{equation}
where we assumed $\bar\rho\neq0$, $\delta\ll1$ and neglected the product of two small quantities, i.e., $\delta\,\nabla\!\cdot \mathbf u$.

The perturbed Euler equation is obtained in the same way
by substituting the relations \eqref{pert-rel} into the Euler equation \eqref{Eul-eq}.
Using the background equation \eqref{Eul-back}
and neglecting all second-order products such as $(\mathbf u \cdot \nabla_x)\mathbf u$,
we obtain:
\begin{equation}\label{Eul-lin}
\dot{\mathbf u} + (H + \gamma_{\alpha})\,\mathbf u
= - \frac{1}{a}\,\nabla_{\!x}\,\phi.
\end{equation}
As seen, equation \eqref{Eul-lin} differs from the standard Newtonian counterpart only
through the additional term $\gamma_{\alpha}(t)$.

To obtain the perturbed Poisson equation, we substitute 
relations \eqref{pert-rel} into equation \eqref{Poi-eq}.
Neglecting all second-order products of the small quantities
$\delta$ and $\phi$ and using the corresponding background relation, we obtain
\begin{equation}\label{Poi-per}
\nabla_{\!x}^{2}\phi = 4 \,\pi \, G a^{2} \bar\rho_{\rm m}\,\delta.
\end{equation}

Finally, eliminating the peculiar velocity $\mathbf u$
and the potential perturbation $\phi$
from the system \eqref{Con-lin}, \eqref{Eul-lin} and \eqref{Poi-per}, we can obtain a differential
equation governing the growth of the density contrast: Taking the divergence of the Euler equation \eqref{Eul-lin} yields:
\begin{equation}\label{div-pert}
\nabla_{\!x}\!\cdot
\left[\dot{\mathbf u} + (H+\gamma_{\alpha})\mathbf u \right]
= - \frac{1}{a}\,\nabla_{\!x}^{2}\phi.
\end{equation}
Then, using the continuity equation \eqref{Con-lin} and the Poisson equation \eqref{Poi-per},
we easily obtain
\begin{equation}\label{dens-gro}
\ddot{\delta}
+ \bigl(2H+\gamma_{\alpha}\bigr)\dot{\delta}
-4\,\pi\, G\,\bar\rho_{m}\,\delta=0 .
\end{equation}

 In the special limit $\alpha=1$,
      equation \eqref{dens-gro} reduces to the standard growth equation for the matter-dominated era in the relativistic and standard Newtonian models \cite{lyth1990evolution}.

\subsection{Solutions of the growth of density inhomogeneities equation: Matter--dominated epoch case study}
\label{Density growth}

In this subsection, using the background solutions of the previous section, we find solutions for equation \eqref{dens-gro}. 

For the matter-dominated era, we insert the background solutions for the density $\bar\rho_m(t)$, the scale factor $a(t)$, and the Hubble parameter $H(t)$ from equations~\eqref{dens-m} and \eqref{mat-a-special} into the growth equation \eqref{dens-gro}. This yields the following evolution equation for the density contrast:
\begin{equation}\label{del-matt}
\ddot{\delta}
+ \left(\frac{3 \,\alpha+1}{3 \,t}\right)\dot{\delta}
+2\left(\frac{3\,\alpha-4}{3\, t^2}\right)\delta=0.
\end{equation}
In the standard case, substituting $\alpha=1$ into \eqref{del-matt}, we obtain
the corresponding relativistic and standard Newtonian counterparts.

\begin{figure}
\centering\includegraphics[width=3in]{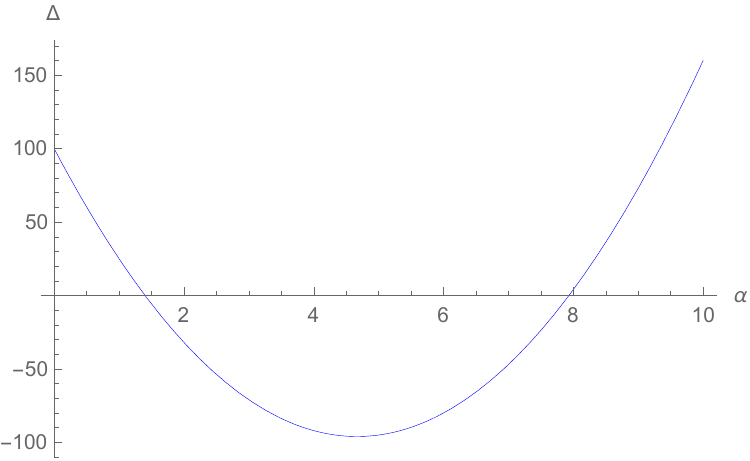}
\centering\includegraphics[width=3in]{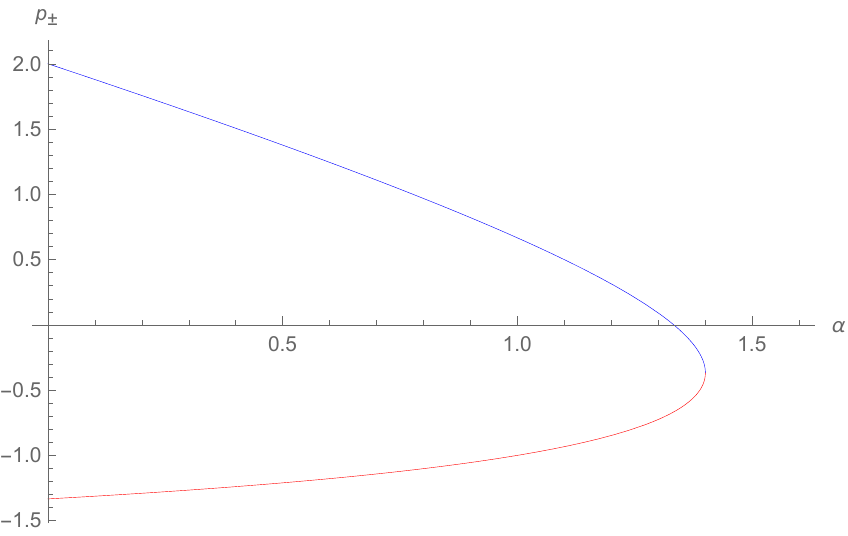}
\caption{Upper panel: The dependence of $\Delta$ on the fractional parameter $\alpha$. Positive values of $\Delta$ correspond to physically allowed (real) solutions for $p_{\pm}$. Lower panel: The evolution of the exponents $p_{+}$ (the blue curve) and $p_{-}$ (the red curve) as functions of $\alpha$, which characterize, respectively, the growing mode for $0<\alpha\lesssim 4/3$ and decreasing modes of the matter density perturbations.
}
\label{p-Delta}
\end{figure}

Equation \eqref{del-matt} admits power–law solutions as $\delta\propto t^{p}$
with the exponents
\begin{equation}\label{p-matter}
p_{\pm}
=\frac{-(3\alpha-2)\pm
  \sqrt{(3\alpha-14)^{2}-96}}{6},
\end{equation}
where $\alpha$ must satisfy the inequality $\Delta\equiv(3\alpha-14)^2-96\ge 0 $. 
Therefore, we obtain two allowed ranges for $\alpha$ as (see also figure \ref{p-Delta})
\begin{equation}\label{real-exp}
\alpha\le\frac{14-\sqrt{96}}{3}\ \approx\ 1.4007
\hspace{2mm}\text{or}\hspace{2mm}
\alpha\ge\frac{14+\sqrt{96}}{3}\ \approx\ 7.9327.
\end{equation}
Since the corresponding standard case (i.e., $\alpha=1$) must also be taken into account, among the above ranges, we will consider the range $0<\alpha\lesssim 1.4007$ as the physically relevant one. Moreover, this constraint ensures that the growth exponents remain real, thereby excluding oscillatory or runaway exponential instabilities.

For the density growth equation \eqref{del-matt}, the general solution can be written as 
\begin{equation}\label{delta-t}
\delta(t)=\delta_+(t_i)\,\left(\frac{t}{t_i}\right)^{p_+}+\delta_-(t_i)\,\left(\frac{t}{t_i}\right)^{p_-}
\end{equation}
where $\delta_+(t_i)$ and $\delta_-(t_i)
$ are the amplitudes of the growing and decreasing modes at the specific initial time $t_i$, respectively, which can be determined by the initial conditions. 

 In the particular case where $\alpha=1$, we get $\delta(t)=\delta_+\,(t/t_i)^{2/3}+\delta_-\,(t/t_i)^{-1}$, which is the same result obtained for the spatially flat FLRW universe filled with a pressure-less fluid \cite{olson1976density}.
 
Moreover, it can be easily shown that $p_-$ is negative for all values considered within the physical range $0<\alpha\lesssim 1.4007$. Therefore, as $\delta^{(d)} \equiv  \delta_-(t_i) (t/t_i)^{p_{-}}$  rapidly dies out as the universe expands and can be neglected at late times, it is interpreted as a decreasing mode. In contrast, $p_+$ is always positive for the values chosen within $0<\alpha\lesssim 4/3$, while it becomes negative when $4/3<\alpha\lesssim 1.4007$, see figure \ref{p-Delta}. Consequently, $\delta^{(g)}\equiv  \delta_+(t_i) (t/t_i)^{p_{+}}$ can be regarded as the growing mode within the range $0<\alpha\lesssim 4/3$, which represents the growth of cosmic structures.

The analysis of the previous section shows that the background solutions corresponding to the radiation-dominated era, the matter-dominated epoch, and the present accelerated phase consistently select a common and observationally admissible range for the fractional parameter $\alpha$. This range is simultaneously supported by theoretical viability conditions and current background-level observational constraints, and is given by \eqref{alpha-obs}.

In this section, we derive an independent upper bound on $\alpha$ by analyzing the growth equation of matter density perturbations. Subsequently, we confront the resulting theoretical predictions with recent observational data related to structure formation. These observations are expected to further restrict the allowed parameter space and to provide a tighter upper bound on $\alpha$, thereby offering a nontrivial consistency check of the fractional framework beyond the background dynamics.

Using equations \eqref{mat-a-special} and \eqref{mat-rho-special}, the relation \eqref{delta-t} can be expressed in terms of the scale factor as
\begin{equation}\label{delta-a}
\delta(a)=\delta_+ \Big(\tfrac{a}{a_i}\Big)^{\frac{3}{2}p_+}+\delta_-\Big(\tfrac{a}{a_i}\Big)^{\frac{3}{2}p_-}.
\end{equation}

We focus on the growing mode $\delta^{(g)}(a)= \Big(\tfrac{a}{a_i}\Big)^{\frac{3}{2}p_+}$ associated with the matter-dominated case from decoupling time, $t_{\mathrm{dec}}$, to the present time, $t_{\mathrm{p}}$, for which we can write
\begin{equation}\label{dec-tr}
\frac{\delta^{\rm(g)}(t_{\rm p},\alpha)}{\delta^{\rm (g)}(t_{\rm dec},\alpha)}
=\Big(\frac{a_{\rm p}}{a_{\rm dec}}\Big)^{\frac{3}{2}p_+}
=\left(\frac{1+z_{\rm dec}}{1+z_{\rm p}}\right)^{\frac{3}{2}p_+},
\end{equation}
where $z$ is the redshift. For the particular case, where $\alpha=1$, using \eqref{p-matter}, the exponent in equation \eqref{dec-tr} reduces to ${\frac{3}{2}p_+}=1$. Therefore, for the standard case, equation \eqref{dec-tr} reduces to  
\begin{equation}\label{dec-tr-standard}
\frac{\delta^{\rm(g)}(t_{\rm p},\alpha=1)}{\delta^{\rm (g)}(t_{\rm dec},\alpha=1)}
=\frac{a_{\rm p}}{a_{\rm dec}}
=\frac{1+z_{\rm dec}}{1+z_{\rm p}}.
\end{equation}

Now, let us calculate the following quantity
\begin{equation}\label{ratio-to-std}
\mathcal{I}\equiv\frac{\delta^{\rm(g)}(t_p,\alpha)}{\delta^{\rm(g)}(t_p,\alpha=1)}
=\frac{\delta^{\rm(g)}(t_p,\alpha)/\delta^{\rm(g)}(t_{\rm dec},\alpha)}{\delta^{\rm(g)}(t_p,\alpha=1)/\delta^{\rm(g)}(t_{\rm dec},\alpha=1)},
\end{equation}
which measures the relative growth of density perturbations with respect to the standard case.  In equation \eqref{ratio-to-std}, we assumed fixed amplitude at decoupling,
\begin{equation}\label{equal-dec}
\delta^{\rm(g)}(t_{\rm dec},\alpha)\simeq \delta^{\rm(g)}(t_{\rm dec}, \alpha=1).
\end{equation}
In the following subsection, we will show that equation \eqref{equal-dec}, which represents a well-justified approximation for our fractional model, can be derived from the Sachs–Wolfe (SW) relation \cite{sachs1967perturbations}.

Now, substituting \eqref{dec-tr} and \eqref{dec-tr-standard} into the quantity \eqref{ratio-to-std}, we obtain
\begin{equation}\label{total}
\mathcal{I}=\left(\frac{1+z_{\rm dec}}{1+z_{\rm p}}\right)^{\frac{3}{2}(p_+-\frac{2}{3})},
\end{equation}

Since the amplitude at decoupling is fixed by the SW normalization,
the quantity $\mathcal{I}(\alpha)$ directly translates into the ratio of present-day clustering amplitudes,
\begin{equation}
\mathcal{I}(\alpha)
= \frac{\sigma_8(\alpha)}{\sigma_8(\alpha=1)},
\label{sigma8_ratio}
\end{equation}
where $\sigma_8$ denotes the root-mean-square mass fluctuation within spheres of radius 
$8\,h^{-1}\mathrm{Mpc}$. For the $\Lambda$CDM cosmology, current observations give 
$\sigma_8(\alpha=1)\simeq 0.81\pm0.006$ \cite{Planck:2018vyg}. 

As a conservative requirement ensuring that structure formation is not excessively suppressed, we impose a lower bound on the relative growth amplitude,
\begin{equation}\label{sigma8_bound}
\frac{\sigma_8(\alpha)}{\sigma_8(\alpha=1)} \gtrsim \mathcal{O}(0.3),
\end{equation}
which provides an observationally motivated constraint on the allowed values of the fractional parameter $\alpha$. Inequality \eqref{sigma8_bound} should not be understood as a direct statistical fit, but rather as a conservative lower threshold
excluding models with unrealistically weak structure growth.

Now, substituting $z_p=0$ and $z_{\rm dec}\simeq1089$ into \eqref{total} and using equations \eqref{sigma8_ratio} and \eqref{sigma8_bound}, we obtain
\begin{equation}\label{Log-p-1}
1090^{(\frac{3}{2}p_{+}-1)}\gtrsim 0.3.
\end{equation}
Taking the natural logarithm of both sides gives
\begin{equation}\label{Log-p-2}
\frac{3}{2}p_{+}-1
\gtrsim
\frac{\ln(0.3)}{\ln(1090)}\simeq -0.172, \qquad
\end{equation}
which yields $p_{+}\gtrsim 0.55$. Substituting the 
analytical expression for $p_{+}(\alpha)$ from \eqref{p-matter} into
this condition gives, on the physically relevant branch near $\alpha=1$, 
$\alpha\lesssim 1.07$.

We see that this upper bound obtained from the observational analysis is slightly smaller than its theoretical counterpart, namely $4/3$ (see figure \ref{p-Delta} for the growing mode). 

Moreover, we can also consider a closer match to the standard amplitude, e.g. $\sigma_8({\alpha})/\sigma_8({\alpha=1})\ge 0.6$, which yields $p_+\gtrsim\ 0.618$, and we therefore obtain $\alpha\ \lesssim 1.03$.  It is seen that a stricter requirement on the present-day clustering amplitude pushes the allowed values of $\alpha$ even closer to the standard value $\alpha=1$. This is also consistent with the broader fractional cosmological framework developed in our previous works, where viable background cosmology and weak-field tests all indicate that the physically relevant regime corresponds to $|\alpha-1|\ll1$ \cite{rasouli2026inflation, rasouli2026emergent}.

\subsection{Sachs--Wolfe effect in the fractional Newtonian model}

On large angular scales, corresponding to super-horizon modes at the time of last scattering, the primary contribution to the cosmic microwave background (CMB) temperature anisotropies is given by the ordinary (non-integrated) SW effect \cite{sachs1967perturbations}. Physically, this effect arises from photons climbing out of gravitational potential wells at the last-scattering surface, leading to a gravitational redshift.
In the simplest and well-justified limit of matter domination at decoupling and adiabatic initial conditions, the temperature anisotropy is given by \cite{sachs1967perturbations}:
\begin{equation}\label{SW-standard}
\frac{\Delta T}{T}
 \;\simeq\; \frac{\phi(\mathbf{r},t)}{3},
\end{equation}
where $\phi$ is the perturbed gravitational potential at decoupling and $\Delta T/T$ represents the temperature fluctuation at large angular
scales. During matter domination and on large scales, the gravitational potential is related to the density contrast 
through the Poisson equation, implying $\phi\propto \delta (t_{dec})$  in the linear regime. 

Equation~\eqref{SW-standard} therefore provides a direct link between the primordial amplitude of density perturbations at recombination and the observed large-angle CMB anisotropies.
Importantly, in the present fractional Newtonian framework, 
the SW relation itself remains unmodified assuming the fractional parameter is close to unity. Therefore, following the standard approach, we assumed that the amplitude of density perturbations at decoupling is fixed by CMB observations and is therefore insensitive to the late-time fractional dynamics. Accordingly, we imposed the approximation
\eqref{equal-dec},
which is well justified since the fractional effects become relevant primarily during the post-recombination evolution. This assumption ensures that any deviation from the standard cosmological scenario originates solely from the modified growth of perturbations at late times.
Concretely, any change in $p_+(\alpha)$ changes the mapping from SW-normalized initial perturbations to present-day clustering observables such as $\sigma_8$ or $S_8$.

It is important to emphasize that the primary 
objective of the present work is not merely to derive an 
observational constraint on the fractional parameter $\alpha$, but 
rather to establish a systematic perturbative framework for 
investigating the evolution of linear matter density perturbations 
within our fractional gravitational model. The observational 
analysis performed here constitutes the first 
phenomenological application of this framework. 
The resulting upper bound on $\alpha$ should therefore 
be interpreted as an independent consistency test of 
the perturbative formalism, complementary to the 
constraints previously obtained from the cosmological 
background evolution \cite{rasouli2026emergent} (including the inflationary epoch \cite{rasouli2026inflation}) 
and weak-field gravitational tests \cite{rasouli2026minimal}. Taken together, these 
independent investigations consistently indicate that the 
physically viable regime remains in the vicinity of the standard limit $\alpha = 1$.


\section{Conclusions and Discussions}
\label{Concl}

In Ref.~\cite{rasouli2026emergent}, we established a minimal fractional modification of the Newtonian gravity, conceptually inspired by gravity on fractal space-time theories or multi-fractional frameworks \cite{calcagni2010fractal,calcagni2010quantum,calcagni2021multifractional}, in which a deformation of the integration measure leads to nonlocal behavior and memory effects in the underlying dynamics. This fractional gravitational framework preserves a well-defined classical limit and remains consistent with general relativity in the weak-field regime, successfully reproducing key experimental tests such as the perihelion precession of Mercury and the gravitational deflection of light \cite{rasouli2026minimal}. At the cosmological level, we have shown that the standard relativistic cosmological equations emerge naturally from this fractional construction \cite{rasouli2026emergent}. As a consequence, the resulting fractional cosmological model provides a unified description of the entire cosmic history at the background level, encompassing pre-inflationary, inflationary, radiation-dominated, matter-dominated, and late-time accelerated phases \cite{rasouli2026emergent,rasouli2026inflation}. Within this framework, all observable quantities acquire a systematic dependence on the fractional parameter $\alpha$, and for its values very close to unity, the model predictions are in full agreement with current observational data.

However, a viable cosmological model has not only been required to produce background solutions consistent with observational data, but it has also been necessary to yield physically acceptable results at the perturbative level, particularly regarding the growth of density inhomogeneities and the formation of large-scale structures.

Moreover, despite extensive investigations in fractional cosmological models, nearly all of these studies have been carried out at the background level. To the best of our knowledge, only a general perturbative analysis, without addressing cosmological solutions and their interpretation, has been performed by us within the framework of the fractional scalar field cosmology \cite{Rasouli:2025qix}. Therefore, there was sufficient motivation to study and interpret first-order perturbations within the framework of the fractional model established in \cite{rasouli2026emergent}.

Therefore, in this work, we have investigated cosmological perturbations during the matter-dominated era, which has played a crucial role since the growth of cosmic structures (such as galaxies and galaxy clusters) has predominantly occurred during this epoch. Any deviation from the standard behavior in this era has therefore left observable imprints that can be directly confronted with cosmological data.

Among the various approaches to perturbing cosmological models, we have adopted the fluid-flow method \cite{lyth1990evolution}. By writing the fractional version of the hydrodynamical continuity, Poisson, and Euler equations corresponding to the special case of the matter-dominated era, we have shown that these equations have been in exact agreement with those of the original fractional model. Moreover, by perturbing these equations, we have derived the corresponding growth equation that governs the matter-dominated epoch and have systematically analyzed its behavior.

We solved the density growth equation for the matter-dominated era. It was observed that both the growth equation itself and its solutions depended explicitly on the fractional parameter $\alpha$, such that in the special case $\alpha = 1$, they reduce to their corresponding counterparts in the relativistic and standard Newtonian cosmological models. Our solution provided a theoretically allowed range for the parameter $\alpha$, see relations \eqref{real-exp}.

Moreover, similar to the aforementioned standard models, two modes, i.e. a growing mode and a decaying mode, were present. In order to ensure consistency with current cosmological observations, we focused on the growing mode. Employing the $\mathrm{SW}$ equation, we derived an observational upper bound on the parameter $\alpha$, cf. Subsection \ref{Density growth}.
This observational upper bound was found to be smaller than the corresponding theoretical upper bound, which had been independently obtained both from the analysis of background solutions in the radiation-dominated, matter-dominated, and present accelerated phases of the universe, and from the theoretical constraints arising from the perturbation analysis, see equation \eqref{alpha-obs}.

Combining the independent constraints arising from background dynamics and perturbation theory, we find that the fractional parameter $\alpha$ is tightly bounded from both sides. The lower bound, $\alpha\gtrsim0.8$, follows from the existence and physical viability of the late-time de Sitter-like background solution, together with the requirement that the effective equation of state parameter $w_{\mathrm{eff}}$ remains within an observationally consistent and physically acceptable range \cite{rasouli2026emergent}. On the other hand, the upper bound is dictated by the dynamics of density perturbations in the matter-dominated era: while theoretical consistency of structure growth imposes $\alpha<4/3$ (see figure \ref{p-Delta} for the growing mode), a significantly stronger observational constraint is obtained using the SW effect on large-scale CMB anisotropies, leading to the bound $\alpha\lesssim1.07$. Remarkably, these lower and upper limits originate from two physically distinct and independent regimes (namely, the background evolution in the de Sitter-like phase and perturbative dynamics in the matter-dominated epoch), thus yielding a robust and non-trivial allowed interval, $0.8 \lesssim\alpha\lesssim 1.07$,
which simultaneously satisfies background consistency, perturbative stability, and observational constraints. We stress that the upper bound $\alpha\lesssim 1.07$ corresponds to a conservative observational criterion. A modestly stricter requirement, e.g. $\sigma_8(\alpha)/\sigma_8(\alpha=1)\gtrsim0.6$, yields the stronger constraint $\alpha\lesssim1.03$, bringing the preferred parameter range even closer to the standard value $\alpha=1$ and reinforcing its consistency with the independent bounds previously derived from background cosmology \cite{rasouli2026inflation} and weak-field gravitational tests \cite{rasouli2026minimal}.

In summary, the preferred range of the fractional parameter obtained here is consistent with the independent constraints previously derived from the cosmological background evolution and weak-field gravitational tests \cite{rasouli2026emergent,rasouli2026inflation,rasouli2026minimal}. This agreement further supports the internal consistency of the proposed fractional framework across different physical regimes.

It is important to emphasize that the analysis of the present work was confined to the linear regime of matter density perturbations on large cosmological scales. The present work was intended as a first perturbative consistency test of the fractional cosmological framework developed in our previous study; see Ref. \cite{rasouli2026emergent}. Accordingly, we did not attempt to describe nonlinear structure formation, halo and subhalo assembly, merger histories, or other stochastic processes associated with strongly nonlinear clustering. 
Therefore, the constraints derived in this work should be interpreted as applying exclusively to the linear growth of large scale matter fluctuations during the matter-dominated era.

Finally, this section is concluded by proposing a systematic framework for deriving the perturbation equations in our emergent $\Lambda$CDM cosmology for fully general cases. 

Since the dynamical equations associated with our fractional cosmology are structurally equivalent to the relativistic cosmological equations \cite{rasouli2026emergent},  the fluid-flow approach can be directly applied to derive the perturbation equations, including the density growth equation. Concretely, by perturbing the quantities $H$, $\rho_{\mathrm{eff}}$, and $p_{\mathrm{eff}}$, we perform a comprehensive and self--consistent perturbative analysis of the fractional model. The perturbation equations obtained by this approach, which are in fact equivalent to the Bardeen and Kodama--Sasaki equations, provide a conceptually transparent and gauge--independent framework for the description of cosmological perturbations. 
Then, using the background solutions obtained in \cite{rasouli2026emergent}, these perturbation equations may in principle be employed to investigate both the radiation-dominated phase and the late-time de Sitter-like regime. However, perturbations in the radiation-dominated era are predominantly oscillatory and do not lead to significant growth of cosmic structures, whereas in the late-time de Sitter-like regime, density perturbations are either frozen or exponentially suppressed. For this reason, the present analysis is restricted to the matter-dominated era, which is the only phase directly relevant to structure formation and to derive observational constraints on the fractional parameter~$\alpha$.
Moreover, such a perturbed formalism may also prove useful for studying perturbations during the early-time inflationary phase. A detailed investigation of the application of this framework to inflationary perturbations will be presented in a forthcoming work.


\begin{acknowledgements}
The author is grateful to the reviewers for their valuable and constructive comments, which significantly improved the clarity of this manuscript. The author acknowledges the FCT grant \textbf{UID/212/2025} Centro de Matem\'{a}tica 
e Aplica\c{c}\~{o}es da Universidade da Beira Interior, as well as
the COST Actions CA23130 (Bridging high and low energies in search of
quantum gravity (BridgeQG)) and CA23115 (Relativistic Quantum Information (RQI)).
\end{acknowledgements}

\bibliographystyle{spphys} 
\bibliography{FracNewRef}

\end{document}